# Multi-Wavelength DFB Laser Based on Sidewall Third Order Four Phase-Shifted Sampled Bragg Grating with Uniform Wavelength Spacing

Xiao Sun[1*], Zhibo Li[1], Yizhe Fan[1], John H. Marsh[1], Anthony E Kelly[1], Stephen. J. Sweeney[1], Lianping Hou[1]

[1]*James Watt School of Engineering, University of Glasgow, Glasgow G12 8QQ, U.K*

**We present the first demonstration of a 1550 nm multi-wavelength distributed feedback (MW-DFB) laser employing a third-order, four-phase-shifted sampled sidewall grating. By utilizing linearly chirped sampled gratings and incorporating multiple true π-phase shifts within the cavity, we achieved and experimentally validated a four-wavelength laser with a channel spacing of 0.4 nm. The device operates stably and uniformly across a wide range of injection currents from 280 mA to 350 mA. The average wavelength spacing was measured at 0.401 nm with a standard deviation of 0.0081 nm. Additionally, we demonstrated a 0.3 nm MW-DFB laser with a seven-channel output, achieving a wavelength spacing of 0.274 nm and a standard deviation of 0.0055 nm. This MW-DFB laser features a ridge waveguide with sidewall gratings, requiring only one metalorganic vapor-phase epitaxy (MOVPE) step and a single III-V material etching process. This streamlined fabrication approach simplifies device manufacturing and is well-suited for dense wavelength division multiplexing (DWDM) systems.**

To address the increasing demand for high-bandwidth density in optical I/O links, integrated dense wavelength division multiplexing (DWDM) systems have been proposed[1, 2], offering advantages such as low power consumption, compact size, and high reliability[3]. A key component of these DWDM systems is a multi-wavelength light source with uniform and stable channel spacing. Various types of multi-wavelength light sources have been proposed, including optical frequency combs[4, 5], mode-locked lasers[6], and multi-wavelength laser arrays[7-9]. Among these, the multi-wavelength laser array is the most widely used, primarily due to its precise control of each laser's lasing wavelength, continuous-wave operation, and high-power output. However, the multi-wavelength laser array requires a large-footprint coupler, such as a multimode interference (MMI) coupler or Y-branch, to combine all laser units into a single waveguide output. This design introduces additional losses and increases the complexity of the fabrication process. A multi-wavelength DFB (MW-DFB) laser with 100 GHz spacing has been proposed [10] based on buried conventional sampled Bragg gratings (C-SBG) grating laser structure which requires twice metalorganic vapor-phase epitaxy (MOVPE) steps for the fabrication.

Reconstruction equivalent-chirp (REC) technology has greatly simplified the challenge of achieving precise wavelength control in laser[11]. Utilizing REC, a uniform-period seed Bragg grating is applied across the entire array, while the wavelength of each laser is determined by sampling the grating. The sampling period dictates the operational wavelength. REC designs usually use C-SBG where 50% of each period is a grating and the remaining 50% has no grating, but C-SBGs only have an effective grating coupling coefficient ($\kappa$) equal to $1/\pi$ of that of a uniform grating [12]. In order to improve the $\kappa$, we have demonstrated several multi-phase shifted sampled Bragg gratings, such as two phase-shifted sampled Bragg grating (2PS-SBG) with an effective $\kappa$ equal to $2/\pi$ of that of a uniform grating [13] and four phase-shifted sampled Bragg grating (4PS-SBG) with an effective $\kappa$ equal to 90% of that of a uniform grating [7, 8].

In this letter, we propose and demonstrate a four-wavelength sidewall MW-DFB laser with 0.4 nm wavelength spacing, employing a third-order, four-phase-shifted sampled Bragg grating (3rd-4PS-SBG) for the first time. Unlike buried C-SBG MW-DFB lasers, which require complex epilayer etching and regrowth, sidewall grating DFB lasers require only a single metalorganic vapor-phase epitaxy (MOVPE) step and one III-V material etching process, significantly simplifying the device fabrication process. The 3rd-order grating mitigates fabrication tolerance issues caused by the reactive ion etching (RIE) lag effect in sidewall grating DFB laser fabrication, while the 4PS design compensates for the reduced $\kappa$ associated with the 3rd-order grating. In the 3rd-4PS-SBG, the effective $\kappa$ value of the -1st channel is notably enhanced, reaching 90% of that of a 3rd-order uniform grating. Additionally, the seed grating period is three times that of a first-order grating, improving the fabrication tolerance of the sidewall grating. Moreover, a multi-wavelength laser with a wavelength spacing of 0.3 nm was also designed and demonstrated, further validating the versatility of the 3rd-4PS-SBG structure across different wavelength systems.

Figure 1(a) depicts the schematic of the 3rd-4PS MW-DFB laser. The designed width of the grating ridge waveguide is 2.5 μm, with a sidewall corrugation depth of 500 nm and a ridge height of 1.92 μm. The grating waveguide has a total length of 2250 μm, and a 100 μm-long curved waveguide is positioned at the end of the grating to minimize facet reflections. The parameter *P* represents the sampling period. Here, REC technology is used to reduce the difficulty of achieving high-precision wavelength control. Based on

REC technology, a seed Bragg grating with a constant period is implemented across the entire array, while the dimensions of the sampling period determine the operational wavelength of each laser. In the 3rd-4PS-SBG structure, each sampling period is evenly divided into four sections, with each adjacent grating section undergoing a π/2 phase shift. The grating period for the ±1st order longitudinal mode of the 3rd-4PS-SBG structure is derived as follows:

$$\frac{1}{\Lambda_{\pm 1}} = \frac{3}{\Lambda_0} \pm \frac{1}{P} \quad (1)$$

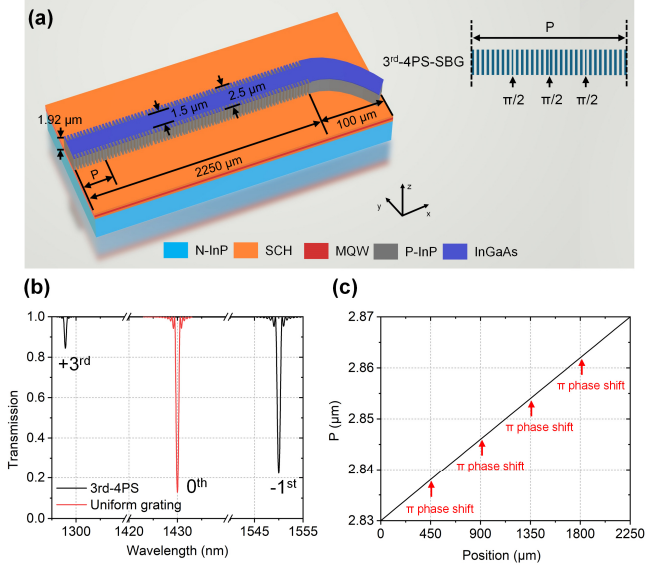

Fig. 1. (a) Schematic of the 3rd-4PS MW-DFB laser. (b) Calculated transmission spectra of the 3rd-order 4PS grating compared with 3rd-order uniform grating. (c) the distribution of the sampling grating period $P$ and the four π phase shifts along the DFB laser cavity.

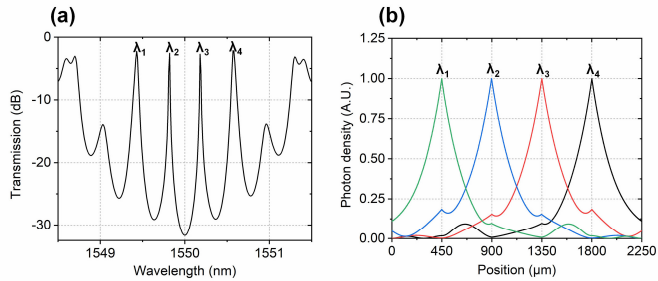

Fig. 2. (a) Calculated transmission spectrum of 3rd-4PS MW-DFB laser, (b) Photon density of the four lasing modes along the DFB laser cavity.

where $\Lambda_{\pm 1}$ refers to the grating period of the ±1st order longitudinal mode, $\Lambda_0$ denotes the seed grating period. Figure 1(b) illustrates the calculated transmission spectra of the 3rd-4PS-SBG structure, indicating that the highest reflection occurs at the –1st longitudinal mode for the third-order seed grating. By chirping the sampling period $P$ as a function of its position within the laser cavity, different lasing wavelengths are achieved. Figure 1(c) shows the distribution of $P$ along the laser cavity, varying linearly from 2.83 to 2.87 µm. To achieve four-longitudinal-mode operation with a wavelength spacing of 0.4 nm, four π-phase shifts are introduced at evenly distributed positions along the grating at 450, 900, 1350, and 1800 µm. The seed grating period is 670.54 nm. Figure 2(a) shows the grating's transmission spectrum, calculated using the coupling-wave method, with four longitudinal mode wavelengths at 1549.4, 1549.8, 1550.2, and 1550.6 nm. Figure 3(b) illustrates the calculated normalized photon density distribution for each longitudinal mode, showing that the photon densities are well-separated, effectively minimizing mode competition.

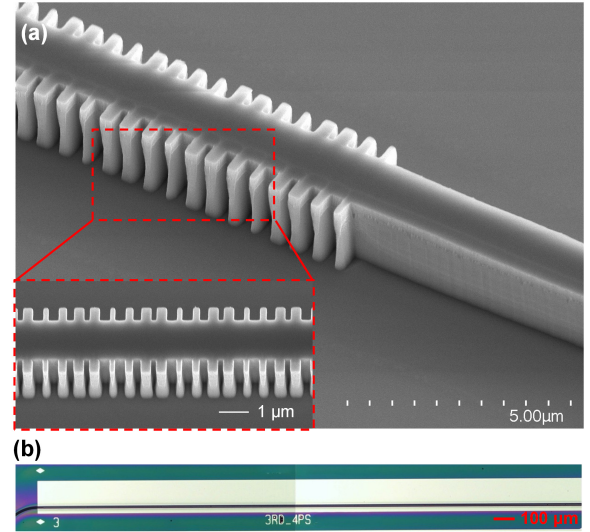

Fig. 3. (a) Side and perspective SEM view of the 3rd-4PS-SBG ridge waveguide, (c) Microscope image of the fabricated MW-DFB laser chip.

The devices were fabricated using an epitaxial structure based on the AlGaInAs/InP material system, which included five quantum wells (QWs) and six quantum barriers, with a QW confinement factor of 5% [14]. The QWs were designed to have a room temperature photoluminescence (PL) wavelength of 1530 nm. The sidewall DFB grating and curved ridge waveguide were patterned onto a 600-nm-thick hydrogen silsesquioxane (HSQ) resist using electron beam lithography (EPBG 5200). The pattern was then transferred to the sample via inductively coupled plasma (ICP) etching with an Oxford Instrument PlasmaPro 300 system, utilizing a $Cl_2/CH_4/H_2$ gas mixture to etch through the upper cladding with a high aspect ratio. Figure 3(a) provides side and perspective scanning electron microscope (SEM) views of the fabricated 3rd-4PS-SBG sidewall ridge waveguide, highlighting the smooth and vertical sidewalls. Subsequent steps, such as the deposition of $SiO_2$ and HSQ passivation layers, $SiO_2$ window opening, P-contact deposition, substrate thinning, and N-contact deposition, followed the same procedures as those used in conventional laser diode fabrication, as detailed in [14]. Figure 3(b) shows the final chip packaged in a box enclosure.

Device testing was conducted under continuous-wave (CW) operation, with the laser chips mounted on a thermoelectric cooler (TEC) set to 20°C. Optical outputs were measured from the uncoated facet on the DFB side using either an optical spectrum analyzer (OSA) or a power meter. Figure 4(a) shows the typical light output and voltage as a function of DFB injection current ($I_{DFB}$). The 2250 μm-long device, featuring a 100 μm-long curved waveguide, has a threshold current of 130 mA. At $I_{DFB}$ = 500 mA, the output power is 7.5 mW with a $V_{DFB}$ = 2 V. Figure 4(b) shows the optical spectrum for $I_{DFB}$ = 290 mA. The lasing wavelengths of the MW-DFB laser are 1558.04, 1558.45, 1558.85, and 1559.24 nm, corresponding to channels CH1 through CH4, respectively.

Table 1. Linear fit results for the lasing wavelengths of the four channels with 0.4 nm spacing at different $I_{DFB}$ values

| $I_{DFB}$ | Wavelength slope | Standard deviation |
|---|---|---|
| 280 mA | 0.403 nm | 0.0081 nm |
| 300 mA | 0.400 nm | 0.0079 nm |
| 320 mA | 0.398 nm | 0.0082 nm |
| 340 mA | 0.401 nm | 0.0083 nm |
| Average | 0.401 nm | 0.0081 nm |

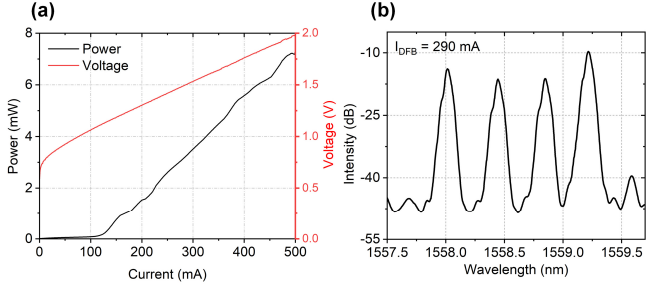

Fig. 4. (a) Typical $P$-$I_{DFB}$ and $V$-$I_{DFB}$ curves for the 0.4 nm spacing four-channel MW-DFB laser, and (b) Measured optical spectrum at $I_{DFB}$=290mA.

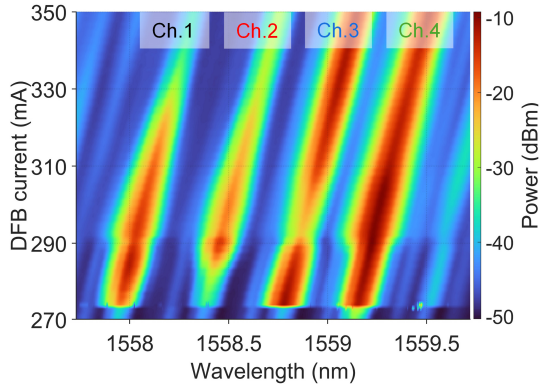

Fig. 5. 2D optical spectra for channels Ch.1 to Ch.4 with $I_{DFB}$ ranging from 270 mA to 350 mA.

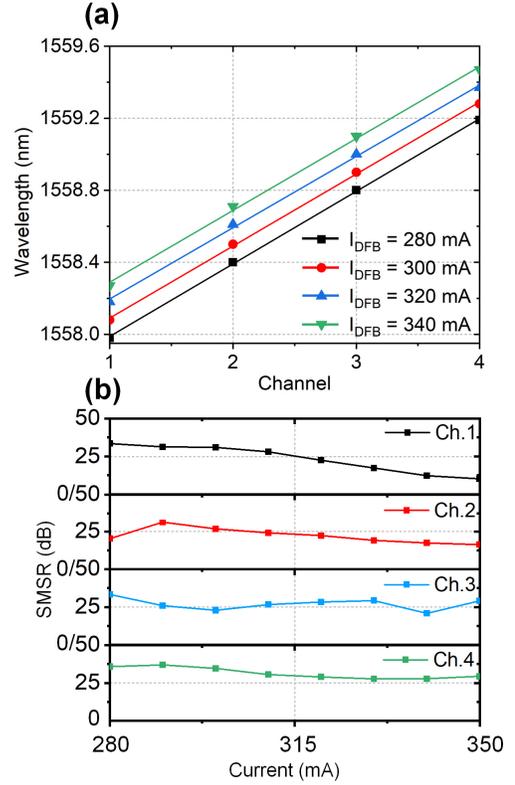

Fig. 6. (a) Lasing wavelengths and their linear fit for the four channels at $I_{DFB}$ = 280 mA, 300 mA, 320 mA, and 340 mA, (b) SMSRs for the four channels as a function of $I_{DFB}$.

Figure 5 presents a two-dimensional (2D) optical spectrum as a function of $I_{DFB}$, showing that the laser maintains stable four-wavelength emission as $I_{DFB}$ varies from 280 mA to 350 mA. The wavelength redshifts with the current are approximately 0.003 nm/mA per wavelength. Figure 6(a) shows the corresponding linear fit to the lasing wavelengths of the four channels in different $I_{DFB}$. The linear fit results are listed in Table 1. The slopes of wavelength are 0.403nm (280 mA), 0.400nm (300 mA), 0.398 nm (320 mA) and 0.401(340 mA). The average wavelength slope is 0.401 nm with a standard deviation of 0.0081 nm. The measured lasing wavelength spacings are nearly identical to the designed target of 0.4 nm. Figure 6(b) shows the side mode suppression ratio (SMSR) of the four devices as a function of $I_{DFB}$. It can be observed that the SMSR for each channel is greater than 15 dB with $I_{DFB}$ ranging from 280 mA to 330 mA.

To further confirm the universality of the proposed structure, the structure with 0.3 nm wavelength spacing is also designed and fabricated. This structure is a 2250 μm-long 3rd-4PS-SGB with seven π phase shifts along the grating. The sampling grating period $P$ varies linearly along the cavity from 5.1 to 5.8 μm, while the seed grating period $\Lambda_{01}$ is 696.7 nm. Figure 7(a) shows the measured $I$-$P$ curve. Seven longitudinal modes are detected at the laser output with $I_{DFB}$ ranging from 280 to 350 mA, as shown in the 2D spectrum map in Figure 7(b). Longitudinal mode competition is more pronounced in this case compared to the four-channel lasers because the grating cavity length per longitudinal mode ($L_{DFB}/7$) is shorter than that for the four-channel lasers ($L_{DFB}/4$), which reduces the $\kappa L$ for each mode. Figure 8 shows the linear fit to the lasing wavelengths for the four channels at different $I_{DFB}$ values. The slopes of the wavelength shifts are 0.272 nm (300 mA), 0.275 nm (310 mA), 0.276 nm (320 mA), and 0.274 nm (330 mA). The average wavelength slope is 0.274 nm with a standard deviation of 0.0055 nm, as detailed in Table 2.

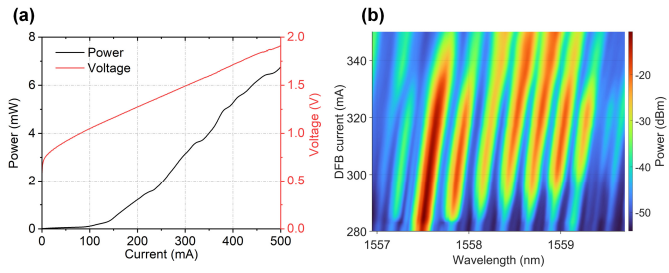

Fig. 7. (a) Typical $P$-$I_{DFB}$ and $V$-$I_{DFB}$ curves for the 0.3 nm spacing seven-channel MW-DFB laser. (b) 2D optical spectra for seven-channel MW-DFB laser with $I_{DFB}$ ranging from 280 mA to 350 mA.

**Table 2. Linear fit results for the lasing wavelengths of the seven channels with 0.3 nm spacing at different $I_{DFB}$ values**

| $I_{DFB}$ | Wavelength slope | Standard deviation |
|---|---|---|
| 300 mA | 0.272 nm | 0.0053 nm |
| 310 mA | 0.275 nm | 0.0050 nm |
| 320 mA | 0.276 nm | 0.0060 nm |
| 330 mA | 0.274 nm | 0.0059 nm |
| Average | 0.274 nm | 0.0055 nm |

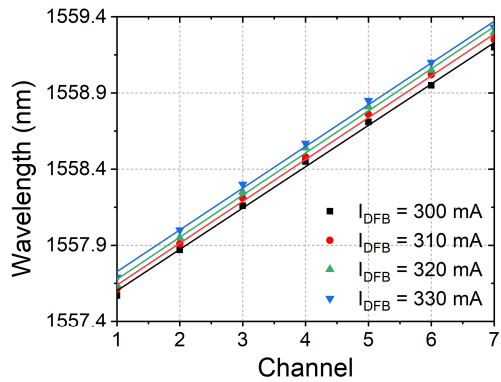

Fig. 8. Lasing wavelengths and their linear fit for the seven channels at $I_{DFB}$ = 300 mA, 310 mA, 320 mA, and 330 mA.

In conclusion, we propose an MW-DFB laser with a 3rd-4PS-SBG, which includes a 3rd-order seed grating, a chirped sampled grating, and multiple π phase shifts evenly distributed along the grating length. This design achieves uniform wavelength spacings of 0.4 nm for four channels and 0.3 nm for seven channels. The standard deviation of the wavelengths is less than 0.01 nm. The SMSR of the 0.4 nm four-channel MW-DFB laser exceeds 15 dB with $I_{DFB}$ ranging from 280 mA to 330 mA. The sidewall grating 4PS-SBG DFB laser array requires only one MOVPE step and one dry etch of the III-V material, simplifying the overall device fabrication process. This streamlined fabrication is particularly advantageous for DWDM sources in price-sensitive applications such as passive optical networks (PONs).

**Funding.** Innovation Funding Service, UK Government (Innovate UK) (10056367)

**Acknowledgment.** We would like to acknowledge the staff of the James Watt Nanofabrication Centre at the University of Glasgow for their help in fabricating the devices.

**Disclosures.** The authors declare no conflict of interest.

**Data Availability Statement (DAS).** Data underlying the results presented in this paper are not publicly available at this time but may be obtained from the authors upon reasonable request.